\documentclass[aps,pre,twocolumn,showpacs,groupeaddress,preprintnumbers,amsmath,amssymb,floatfix]{revtex4-1}
\usepackage{graphicx,bm,amsmath,amssymb,natbib,url,epsfig}
\usepackage{dcolumn}
\usepackage{hyperref}

\usepackage{color}
\definecolor{darkgreen}{rgb}{0.1,.6,.1}
\definecolor{greenblue}{rgb}{0.0,.1,.4}

\usepackage[normalem]{ulem}

\begin{document}

\title{Imperfect traveling chimera states induced by local synaptic gradient coupling} 

\author{Bidesh K. Bera${}^{1}$}
\email{bideshbera18@gmail.com}
\author{Dibakar Ghosh${}^{1}$}
\email{diba.ghosh@gmail.com}
\author{Tanmoy Banerjee${}^{2}$}
\email{tbanerjee@phys.buruniv.ac.in}
\affiliation{%
${}^{1}$ Physics and Applied Mathematics Unit, Indian Statistical Institute, Kolkata-700 108, India.\\
${}^2$Department of Physics, University of Burdwan, Burdwan 713 104, West Bengal, India.
}%

\received{:to be included by reviewer}
\date{\today}

\begin{abstract}
In this paper we report the occurrence of chimera patterns in a network of neuronal oscillators, which are coupled through {\it local}, synaptic {\it gradient} coupling. We discover a new chimera pattern, namely the {\it imperfect traveling chimera} where the incoherent traveling domain spreads into the coherent domain of the network. Remarkably, we also find that chimera states arise even for {\it one-way} local coupling, which is in contrast to the earlier belief that only nonlocal, global or nearest neighbor local coupling can give rise to chimera; this find further relaxes the essential connectivity requirement of getting a chimera state. We choose a network of identical bursting Hindmarsh-Rose neuronal oscillators and show that depending upon the relative strength of the synaptic and gradient coupling several chimera patterns emerge. We map all the spatiotemporal behaviors in parameter space and identify the transitions among several chimera patterns, in-phase synchronized state and global amplitude death state.
\end{abstract}


\maketitle

\section{Introduction}
\label{sec:intro}

The research on the collective behavior of {\it identical} oscillators has been revitalized with the discovery of chimera \cite{chireview}. Kuramoto and Battogtokh \cite{KuBa02} discovered that the population of identical oscillators is subdivided into two incongruous domains: In one domain the neighboring oscillators are synchronized whereas in another domain the oscillators are desynchronized. This intriguing spatiotemporal state was named as chimera by Strogatz \cite{st1} and it has been in the center of attention over the last decade.  After its discovery in phase oscillators under nonlocal coupling by Kuramoto and Battogtokh \cite{KuBa02} other chimera states have also been discovered. Amplitude mediated chimera was reported in \cite{amc_sethia} where, under strong {\it global coupling}, a network of complex Ginzberg-Landau oscillators show fluctuations in both the phase and amplitude part in the incoherent region. Later, {\it amplitude chimera} was discovered by \citet{scholl_CD} where all the oscillators in a network have the same phase velocity but differ in amplitude in the incoherent region. 
Recently, chimera state is observed in {\it local} nearest neighbor coupling \cite{g,diba15} also, which proves that a nonlocal or strong global coupling is not an essential requirement to achieve chimera.

The robustness of chimera has been proved through a series of experiments: The first experimental observation of chimera was reported in optical system \cite{HaMuRo12} and chemical oscillators \cite{TiNkSh12}. Later, chimera has been observed experimentally in several other systems; for example, in mechanical system \cite{MaThFo13} and time-delay laser system \cite{LaPeMa15} to name a few. Recently, \citet{raj_minimal} reports the experimental observation of chimera state in globally coupled minimal network of four optoelectronic oscillators. Apart from these man-made experimental setups the occurrence of chimera has been reported in real physical and biological systems: For example in SQUID meta material \cite{expt_meta,expt_meta2}, ecology \cite{DuBa15,scholleco}, and quantum system \cite{schoell_qm}.

Ever since its discovery, the chimera state has been strongly connected to the several neuronal processes, like unihemispheric slow-wave-sleep of some mammals in aquatic species, like dolphin, eared seals and manatees \cite{uni,uni2}, and some migratory birds \cite{uni2}. During slow-wave-sleep, these species shut-down only one cerebral hemisphere of the brain, and close the opposite eye. During this time, the other half of the brain monitors what is going in the environment (for migratory birds) and controls breathing functions (for aquatic mammals). It strongly indicates that in the awake part of the cerebral hemisphere, neuronal oscillators are desynchronized while in the sleepy part neuronal oscillators are very much synchronized, which resemblances the chimera state.

Therefore, systematic studies on chimera in neuronal systems deserve special attention. Earlier works on network of neuronal oscillators considered FitzHaugh-Nagumo model having type-II excitability \cite{fhn-prl,fhn-anna} and SNIPER model of type-I excitability \cite{type1-ph}; those works deal with the spiking neurons (as those models could not produce bursting behavior). In this context neuronal oscillation modeled by the Hindmarsh-Rose (HR) model is much more realistic because, depending upon parameter values it shows both type-I and type-II excitability and can produce a plethora of physiologically relevant neuronal oscillations, like square wave bursting (both periodic and chaotic), plateau bursting, spiking, mixed mode bursting, etc. \citet{hr-ijbc} reported the occurrence of chimera in a network of HR-neuron under nonlocal coupling. But the considered nonlocal coupling has an ideal rectangular kernel, which has less biological relevance as per as the coupling function is concerned.

In this context, synaptic coupling is one of the most realistic coupling schemes through which neurons are connected. In the context of inter-neuronal communication two variants of synapses exist: Chemical and electrical synapse.
\citet{diba15} considered the chemical synaptic coupling with nonlocal, global and local nearest neighbor coupling topology and established the occurrence of chimera. But in that work, only excitatory coupling with symmetric synaptic coupling strength was considered in all three coupling topologies. Normally, in a network of neuronal oscillators both excitatory and inhibitory interactions are present \cite{iz-book}. In the absence of time delay, in general, fast excitatory coupling leads to synchronization and fast inhibitory coupling promotes anti synchronization. \citet{igor1} showed that the  synergistic effect of excitation and inhibition leads to complete synchronization (i.e., synchrony in both burst and spike) with lesser coupling threshold. Thus, it is important to study their simultaneous effect on the occurrence of multi-cluster synchrony or chimera state. The simultaneous effect of excitatory and inhibitory coupling is best represented by {\it gradient coupling} : in this coupling scheme, one can control the strength of excitation and inhibition by simply controlling the parameter associated with the gradient coupling. Earlier the effect of the gradient coupling was studied in the context of oscillation suppression in nonlinear oscillators with time-delay diffusive coupling \cite{grad-ad}, synchronization of nonlinear coupled systems \cite{grad-syn} and control of spatiotemporal chaotic systems \cite{grad-cont,grad-cont2}, but its influence on the occurrence of chimera has not been studied yet.

In this paper, for the first time, we investigate the effect of local chemical {\it synaptic gradient} coupling on the occurrence of chimera state. We choose a network of identical square-wave bursting Hindmarsh-Rose neuronal oscillators and show that depending upon the proper choice of the coupling parameter associated with the gradient coupling we can attain three different coupling topologies, namely {\it one-way} excitatory local coupling, nearest neighbor {\it asymmetric} excitatory coupling and {\it simultaneous excitatory-inhibitory} coupling. Remarkably, we find chimera even in the {\it one-way} excitatory local coupling that further relaxes the essential requirement of getting chimera in a network. We identify and confirm the occurrence of a new chimera pattern, namely the {\it imperfect traveling chimera} in the network with nearest neighbor asymmetric excitatory coupling. We further map in parameter space the transitions among chimera patterns, synchronized state, and amplitude death state with the variation of the synaptic and gradient coupling strengths.

\begin{figure}
\centering
\includegraphics[width=0.4\textwidth]{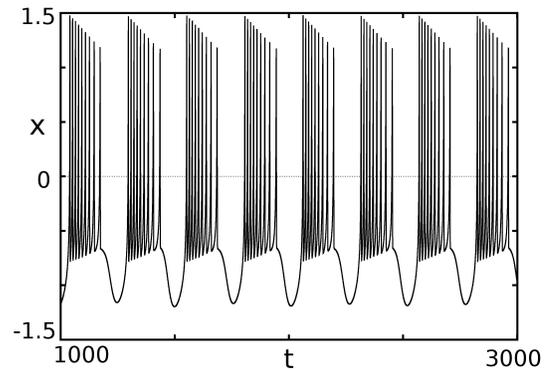}
\caption{\label{F:x} (Color online) Square wave bursting: time series of $x$ of an individual HR neuron given by Eq.~\ref{eq1} ( $a=2.8$, $\alpha=1.6$, $c=5$, $b=9$, $\mu=0.001$).}
\end{figure}
\section{Coupling scheme}
We consider a network of $N$ identical Hindmarsh-Rose neurons \cite{hr,igor3} coupled via a local synaptic {\it gradient} coupling given by

\begin{subequations}\label{eq1}
\begin{align}
\dot{x_i}&=ax_i^2-x_{i}^{3}-y_i-z_i\nonumber\\
&
+(V_s-x_i)\left((\epsilon+r)\Gamma(x_{i+1})+(\epsilon-r)\Gamma(x_{i-1})\right),\\
\dot{y_i}&=(a+\alpha)x_{i}^{2}-y_i,\\
\dot{z_i}&=\mu(bx_i+c-z_i).
\end{align}
\end{subequations}
$i=1,2,\cdots, N (N\ge3)$. The variable $x_i$ represents the membrane potential of the $i$-th HR neuron, $y_i$ is the fast current (associated with $\mbox{Na}^+$ or $\mbox{K}^+$), and $z_i$ represents the slow current (associated with $\mbox{Ca}^{2+}$). The parameter $\mu$ determines the ratio of slow / fast time scales. The sigmoidal function $\Gamma(x_i)=1/[1+\mbox{exp}(-\lambda(x_i-\Theta_s))]$ represents the fast threshold modulation synaptic coupling \cite{igor2} where $\lambda$ determines the slope of the function and $\Theta_s$ is the firing threshold. We use the periodic boundary condition $(x_{N+1}=x_1)$. We take the reversal potential $V_s=2$ such that $V_s>x_i(t)$ is always satisfied; also, we choose the synaptic threshold $\Theta_s=-0.25$ to ensure that every spike in a burst can reach the threshold value \cite{igor1,igor2} and the slope of the sigmoidal function $\lambda=10.$

The most interesting and important part of the coupling function of \eqref{eq1} is the gradient coupling: if we choose $r=0$, then Eq.~\ref{eq1} represents a synaptic nearest-neighbor {\it excitatory} coupling with the synaptic coupling strength $\epsilon~(>0)$. But, for $r>0$, depending upon the relative values of $r$ and $\epsilon$ three distinct coupling scenarios can be arose:

CASE-I ~~$\epsilon>r$: Here we have excitatory coupling [since, $(\epsilon\pm r)>0$ ], but now the $i$-th neuron is coupled to its nearest neighbors with two different synaptic coupling strengths; this introduces {\it asymmetry} in the coupling.

CASE-II~~ $\epsilon=r$: This condition represents one-way local excitatory coupling with the effective synaptic coupling strength $2\epsilon$. As $V_s>x_i(t)$ for all time $t$ and $x_i(t)$, the coupling is positive and synapse is excitatory, i.e., the input from the $(i+1)$-th neuron to the $i$-th neuron can enhance the activity of this neuron.

CASE-III ~~$\epsilon<r$: Now the $i$-th neuron is connected to the $(i+1)$-th neuron through an {\it excitatory} coupling with an effective synaptic coupling strength $(\epsilon+r)$, but it is connected to the $(i-1)$-th neuron via an {\it inhibitory} coupling with an effective synaptic coupling strength $(\epsilon-r)$. In this case the input to the $i$th neuron from the $(i+1)$-th neuron via synaptic coupling can enhance its activity and from the $(i-1)$-th neuron via synaptic coupling can suppress its activity.  Thus, this condition leads to simultaneous occurrence of excitation-inhibition in the HR-neuron network, which is the most important case in the context of neuroscience. Therefore, we denote $r (>0)$, which governs the strength as well as nature of the coupling, as the {\it gradient coupling strength} and $\epsilon (>0)$ as the {\it synaptic} coupling strength.

In the following we will explore the dynamics of the network under these three distinct coupling conditions; our main emphasis will be to identify the parameter region of the synaptic and gradient coupling strengths where chimera occurs.

\begin{figure}
\centering
\includegraphics[width=0.49\textwidth]{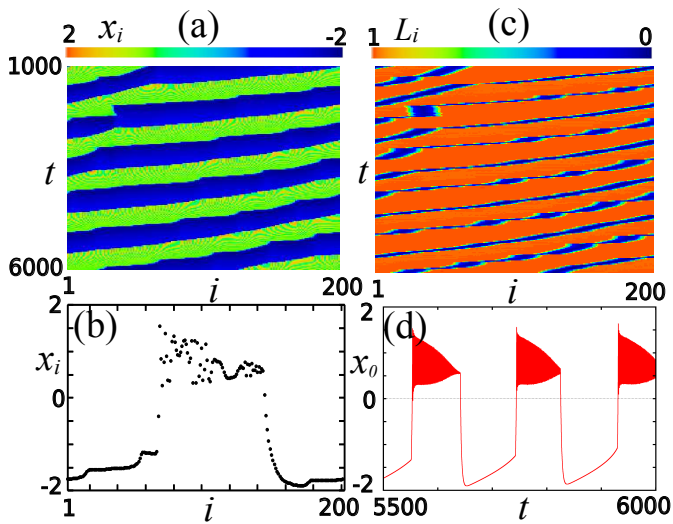}
\caption{\label{F:egr1} (Color online)  $\epsilon>r$: Imperfect traveling chimera state of an ensemble of HR-neurons with asymmetric excitatory local coupling. (a) Spatiotemporal plot as a function of the neuron index $i=1, 2, \cdot \cdot \cdot, N=200$ for $r=0.2$ and $\epsilon=0.7$ ($\epsilon>r$). (b) Snapshot of the variable $x_i$ at $t=2000$, (c) local order parameter with node index $i$. Red (gray) indicates coherent and blue (dark gray) represents incoherent domains. (d) Time series of an individual neuron $x_0$.}
\end{figure}
\begin{figure}
\centering
\includegraphics[width=0.49\textwidth]{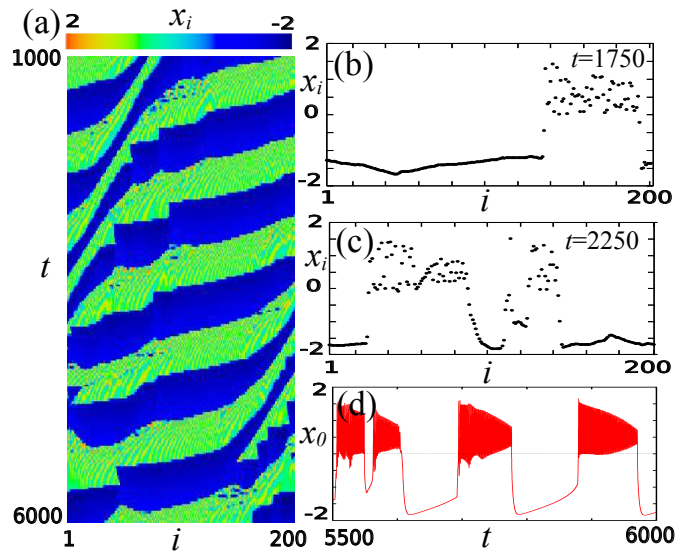}
\caption{\label{F:egr2} (Color online)  $r=0.2$ and $\epsilon=0.56$ ($\epsilon>r$): Imperfect traveling chimera with increasing imperfection (a) Spatiotemporal plot as a function of the neuron index $i=1, 2, \cdot \cdot \cdot, N=200$. Snapshot of the variable $x_i$ at (b) $t=1750$ showing {\it one-headed chimera}, (c) $t=2250$ showing {\it two-headed chimera}. (d) Time series of an individual neuron $x_0$.}
\end{figure}

\section{Results}
We explore the spatiotemporal dynamics of the considered network under three different relative values of $\epsilon$ and $r$. We consider $N=200$ and integrate Eq.~\ref{eq1} using the fourth-order Runge-Kutta algorithm (step size=0.001). Throughout the paper we use the following system parameters of HR-neuron: $a=2.8$, $\alpha=1.6$, $c=5$, $b=9$, $\mu=0.001$. With this system parameters an isolated HR-neuron shows square wave bursting (see Fig.~\ref{F:x}).

\subsection{CASE-I: $\epsilon>r$, Asymmetric excitatory coupling: Imperfect traveling chimera }
At first we start with the condition $\epsilon>r$. In this condition, the $i$-th node is connected to $(i+1)$-th node with an effective synaptic coupling strength $(\epsilon+r)$, whereas it is connected to the $(i-1)$-th node with  an effective synaptic coupling strength $(\epsilon-r)$. Since $\epsilon>r$, thus the coupling is always excitatory in nature with asymmetric coupling strengths.

We fix the value of gradient coupling strength at $r=0.2$ and vary the synaptic coupling strength $\epsilon$. For lower value of $\epsilon$ ($\epsilon\lesssim0.5$), all the neurons show a disordered turbulent  state. With increasing $\epsilon$ we observe that the coupled system enters into a state where certain neurons are synchronized but the remaining neurons are in an incoherent state, which is a signature of chimera state. But this chimera state is not static in the sense that it travels with time in the spatiotemporal domain. Significantly, this state is not a pure traveling chimera as during its travel some synchronized neurons are added to the traveling incoherent state. This type of traveling chimera where the incoherent domain spreads to the synchronized domain is a {\it new observation} and we call it an {\it imperfect traveling chimera state}. Earlier, Refs.~\cite{impchimera2,impchimera} reported the existence of imperfect chimera in coupled oscillators and phase oscillators with inertia, but the {\it imperfect traveling chimera state} has not been reported previously. The scenario is shown in Fig.~\ref{F:egr1}: here we take an exemplary value $\epsilon=0.7$ and $r=0.2$ (note that $\epsilon>r$).  Figure~\ref{F:egr1} (a) shows the time evolution of the membrane potential of all neurons for a long time run. This spatiotemporal plot clearly shows that the spatiotemporal patterns of varying width traveling in space and time. Fig.~\ref{F:egr1}(b) shows the snapshot of all the neurons at time $t=2000$. The spatial coexistence of coherent and incoherent neurons can be clearly observed from the snapshot. 

To characterize the coherence-incoherence pattern and chimera we use the notion of  {\it local order parameter} (note that, since here we have traveling chimera, thus, the mean phase velocity is not an appropriate measure). The local order parameter  actually represents the local ordering of the oscillators and thus the degree of (in)coherency; it is defined as \cite{scholleco,local-order1}
\begin{equation}
L_i=\left|\frac{1}{2\delta}\sum_{|i-k|\le\delta}e^{j\Phi_k}\right|
\end{equation}
 where $j=\sqrt{-1}, i=1, 2, \cdots, N $ and $\delta$ is the nearest neighbors on both sides of the $i$-th oscillator. Here, we define $\Phi_i=\mbox{arctan}(y_i/x_i)$ as the geometric phase of the $i$-th HR-neuron, which is a good approximation as long as $\mu$ is small ($<<1$). The local order parameter of the $i$-th neuron, $L_i\approx1$, indicates that the $i$-th neuron belongs to the coherent part of the chimera state, i.e., $L_i=1$ means the maximum ordering or coherency. On the other hand, $L_i\approx0$ represents that the $i$-th neuron belongs to the incoherent neighboring nodes. We take the window size of spatial average as $\delta=12$ and compute the local order parameter $L_i$ of each neuron for long time interval, which is shown in Fig.~\ref{F:egr1}(c). Red(grey) region represents the coherent nodes and in between two consecutive traveling coherent domains, we have incoherent traveling domains [represented in blue (dark gray)]. It is important to note that the width of the incoherent blue (dark gray) domain changes as they travel in space: this indicates the spreading of the incoherent domain to the synchronized domain, which is the clear signature of {\it imperfect traveling chimera state}. The local dynamics, i.e., dynamics of each neuron in this chimera state are identical and a typical time series of a single HR neuron is shown in Fig.~\ref{F:egr1}(d), which shows that individual neurons are in {\it plateau bursting} mode.

For a very narrow region in parameter space we observe that the ``imperfectness" of the imperfect traveling chimera increases for slightly lower value of $\epsilon$ in comparison with the $\epsilon$-value for which  imperfect traveling chimera occurs. Fig.~\ref{F:egr2} shows this case for $\epsilon=0.56$ ($r=0.2$). Here we can observe that the chimera travels erratically in space and time; also, new domain is created (destroyed) from (to) a completely different type of domain [see Fig.~\ref{F:egr2}(a)]. This is also clearly seen in Figs.~\ref{F:egr2}(b) and \ref{F:egr2}(c): here Fig.~\ref{F:egr2}(b) shows the snapshot at $t=1750$ that shows a one-headed chimera, but at $t=2250$ a two-headed chimera is created, which is shown in Fig.~\ref{F:egr2}(c). Interestingly, the  behavior of the individual neurons at this state shows a mixed mode oscillation of square-wave bursting and plateau bursting, which is shown in Fig.~\ref{F:egr2}(d).

\begin{figure}
\centering
\includegraphics[width=0.45\textwidth]{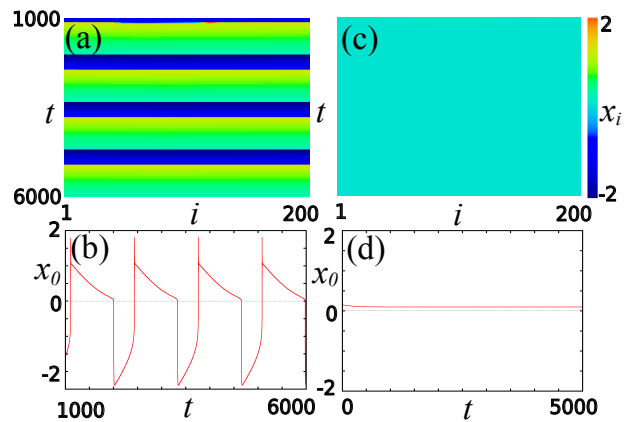}
\caption{\label{F:egr3} (Color online)  $\epsilon>r$, $r=0.2$ :(a) $\epsilon=1.3$, globally synchronized populations of neurons. (b) The corresponding time series of a single neuron. (c) $\epsilon=1.6$, global amplitude death. (d) The corresponding time series of a single neuron.}
\end{figure}

Apart from chimera patterns we also investigate the collective behavior of the network at larger values of $\epsilon$. For an increasing $\epsilon$ (fixed $r$), beyond imperfect traveling chimera the network shows global synchrony. This is shown in Fig.~\ref{F:egr3}(a) for $\epsilon=1.3$. Note that here individual neurons show plateau bursting [Fig.~\ref{F:egr3}(b)]. Further increase in $\epsilon$ results in global amplitude death [Fig.~\ref{F:egr3}(c)] where the oscillations of all the neurons cease. Fig.~\ref{F:egr3}(d) shows the time series of an individual neurons. The transition from oscillatory state to amplitude death state has a broad relevance in neuroscience where we can control the neuronal outputs using this local chemical synaptic coupling.

At this point it is interesting to note the change in the time series with increasing $\epsilon$. For the asynchronous state the HR-neurons show square-wave bursting (not shown) but with increasing $\epsilon$ the square-wave bursting changes into the plateau bursting [Fig.\ref{F:egr1} (d)] (via the disappearance of homoclinic bifurcation \cite{igor1}), through a mixed time series of square-wave and plateau bursting [Fig.\ref{F:egr2} (d)]. In the synchronized state neurons show a pure plateau bursting which agrees with the observation of \citet{igor2} that plateau bursting promotes synchrony. The frequency of plateau bursting decreases with increasing $\epsilon$  it eventually reaches an infinite period, i.e., now global amplitude death emerges [Fig.\ref{F:egr3} (d)]: this typical pattern is indicative that amplitude death appears through a saddle-node bifurcation \cite{dhamala}. 


\begin{figure*}
\centering
\includegraphics[width=0.75\textwidth]{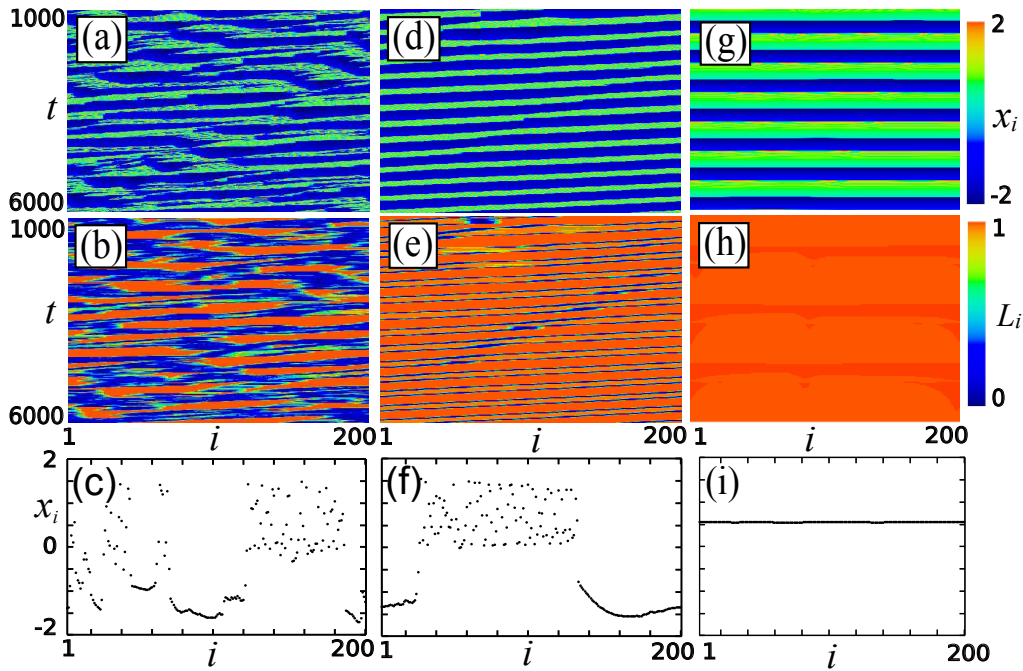}
\caption{\label{F:er} (Color online)  One-way excitatory coupling $\epsilon = r$: (a-c) $\epsilon = r=0.56$: imperfect chimera; (a) spatiotemporal plot (b) Local order parameter and (c) snapshot at $t=3500$ shows multi-headed chimera. (d-f) $\epsilon = r=0.6$: traveling chimera; (d) spatiotemporal plot (e) Local order parameter and (f) snapshot at $t=3500$ shows one-headed chimera. (g-i) $\epsilon = r=0.7$: Synchronized state; (g) spatiotemporal plot (h) Local order parameter and (i) snapshot at $t=3500$ shows synchronized $x_i$.}
\end{figure*}

\subsection{CASE-II: $\epsilon=r$, one-way excitatory coupling: Imperfect chimera and traveling chimera}
As we discussed earlier for $\epsilon=r$, each $i$-th neuron is connected to the $(i+1)$-th neuron with an effective synaptic coupling strength of $2\epsilon$. If we increase $\epsilon$ from a lower value we observe complete turbulence up to  $\epsilon\approx0.5$. We find two chimera patterns beyond that: one is {\it imperfect chimera} that occurs for a small range of $\epsilon$ and beyond that {\it traveling chimera}. Figs.~\ref{F:er}(a)-(c) show the spatiotemporal plot of amplitude and local order parameter ($L_i$), and snapshot at a time instant ($t=3500$), respectively for the imperfect chimera at $\epsilon=0.56$. The snapshot in Fig.~\ref{F:er}(c) shows the multi-headed chimera pattern. The local order parameter [Fig.~\ref{F:er}(b)] suggests that the incoherent region [blue (dark gray)] is not static in space and time, it rather changes erratically indicating an imperfect chimera. Earlier, the imperfect chimera was also observed in Kuramoto model with inertia using nonlocal coupling \cite{impchimera}.

If we increase the value of synaptic coupling strength $\epsilon$, we observe traveling chimera. Figures~\ref{F:er}(d)-(f) show the traveling chimera state at $\epsilon=0.6$. The local order parameter [Fig.~\ref{F:er}(e)] shows that, indeed, the incoherent (and the coherent) domain is not static in time and space. From the snapshot at $t=3500$ [Fig.~\ref{F:er}(f)], it is observed that the system depicts a one-headed chimera state.  Further increases in $\epsilon$ results in a globally synchronized state. Figures~\ref{F:er}(g)-(i) depict this state for $\epsilon=0.7$. Here the local parameter [Fig.~\ref{F:er}(h)] attains a value $L_i\approx1$, which indicates the presence of complete synchrony among the neurons. The snapshot at $t=3500$ of $x_i$ [Fig.~\ref{F:er}(i)] supports this finding. Note that for $\epsilon=r$ synchronization occurs at a much lower value in comparison with that of $\epsilon>r$. Finally, for $\epsilon\gtrsim 1.45$ global amplitude death occurs where all the neurons come to a common steady state and achieve a quiescent state (not shown in the figure).

It is important to note that earlier chimera state has been found under non local, global and nearest neighbor local coupling. Here, the occurrence of the chimera state even for $\epsilon=r$ actually reduces the essential requirement of coupling function to  {\it one-way nearest neighbor coupling}. This new finding may be extended to other coupling scheme also (i.e., other than chemical synaptic coupling). Previously, the existence of chimera states also reported in reaction-diffusion system where each systems are locally coupled by diffusion \cite{g,li}. Although, it is noted that such form of local coupling through diffusion actually gives nonlocal effect due to the fast variable. In this way a three components reaction-diffusion model is reduced to an effective two components model where the third component changes so fast that it can be eliminated adiabatically and it is represented by a nonlocal coupling \cite{shima}.

\begin{figure}
\centering
\includegraphics[width=0.49\textwidth]{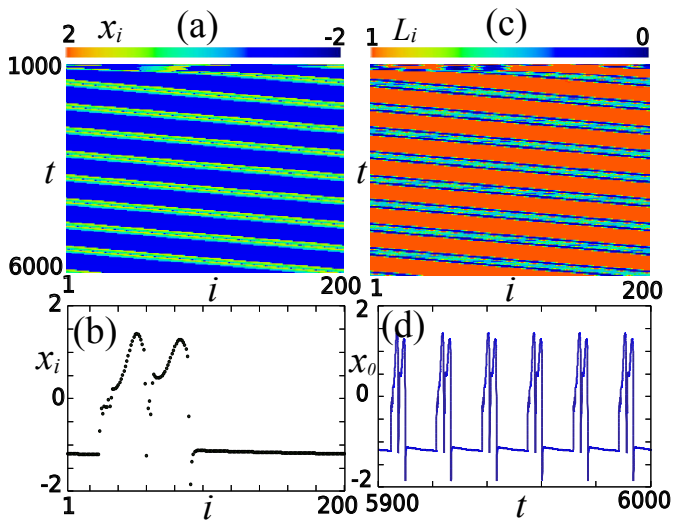}
\caption{\label{F:elr1} (Color online) $\epsilon<r:$ Traveling chimera state in an ensemble of HR-neuron with excitation-inhibition local coupling  $r=8$ and $\epsilon=0.6$. (a) Spatiotemporal plot as a function of node index $i=1, 2, \cdot \cdot \cdot , N=200$ (b) Snapshot of the variable $x_i$ at $t=3300$. (c) Local order parameter with node index. Red (gray) indicates coherent and blue represents incoherent domains. (d) Time series of an individual neuron.}
\end{figure}

\begin{figure}
\centering
\includegraphics[width=0.49\textwidth]{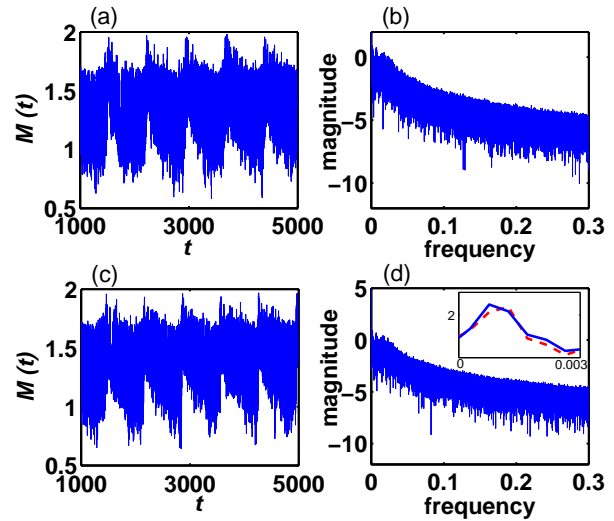}
\caption{\label{power} (Color online) Simultaneous presence of excitatory and inhibitory coupling for $\epsilon<r$: variation of maximum values of $x_i(t)$ with time $t$, for (a) $r=8.0$ and (c) $r=6.0$.  Corresponding Fourier transform of $M(t)$ for (b) $r=8.0$ and (d) $r=6.0$.  The inset in (d) shows the maximum amplitudes occur for $r=8.0$ (red dashed line) and $r=6.0$ (blue color line).  }
\end{figure}
\subsection{CASE-III: $\epsilon<r$, Simultaneous excitation-inhibition: Traveling chimera}
We now explore the spatiotemporal dynamics of the considered network for $\epsilon<r$. Under this condition a node is connected to its nearest right neighbor through an excitatory coupling whereas to its left nearest neighbor through an inhibitory coupling. Earlier \citet{igor1} showed that the simultaneous presence of excitation and inhibition provides a synergistic effect to improve (reduce) the synchronization threshold \cite{igor1}. They show that the synchronization threshold is reduced due to the fact that the simultaneous presence of excitation and inhibition turns the dynamics from square-wave bursting to plateau bursting. Here also, we investigate the effect of simultaneous presence of excitation and inhibition on the occurrence of chimera state.

We choose $\epsilon=0.6$ and increase $r$ from $0.6$ to higher values. We observe traveling chimera with increasing  $r$. For  larger values of $r$ traveling chimera state gets weaker. Fig.~\ref{F:elr1}(a) shows the spatiotemporal plot of the traveling chimera for an exemplary value $r=8.0$. The corresponding snapshots of $x_i$ at $t=3300$ is shown in Fig.~\ref{F:elr1}(b). It can be observed that in the ``bump-like" spatial domains not all the nodes are incoherent but only certain number of nodes are incoherent whereas others maintain a phase-locked state. Note that the direction of the traveling chimera for $\epsilon<r$ is just the {\it reverse} to that for $\epsilon=r$ [Fig.~\ref{F:er}(d)] or the imperfect chimera of $\epsilon>r$ [Fig.~\ref{F:egr1}(a)].  Fig.~\ref{F:elr1}(c) demonstrates the time evolution of the local order parameter $L_i$ of all the neurons showing strips of coherent and incoherent nodes that are traveled with time. Red and blue region represent the coherent and incoherent domains, respectively. It is also noteworthy to track the change in the time series with increasing $r$: for lower $r$, the individual neuron shows square-wave bursting, but with the increase of $r$ bursting changes into multiple spiking, and for a large $r$, time series eventually shows a single spike. Fig.~\ref{F:elr1}(d) shows the time series corresponding to traveling chimera, which is a train of two pulses.

Next, we investigate whether {\it inhibitory} coupling alone can give rise to chimera. For this we set $\epsilon=0$ and vary $r$:  we get incoherent state for higher values of $r$. Significantly, we find that, although excitatory coupling can induce several chimera patterns, the inhibitory coupling alone can not induce chimera state. But, the presence of inhibitory coupling along with the excitatory coupling supports chimera and traveling chimera. It is also noted that for $\epsilon<0.5$, complete turbulence occurs irrespective the value of $r$; thus chimera and synchronization states (and also the amplitude death state) depend largely upon $\epsilon$: only beyond a certain value of $\epsilon$, the presence of $r$ makes the system to show several chimera patterns.

\par
\begin{figure}
\centering
\includegraphics[width=0.45\textwidth]{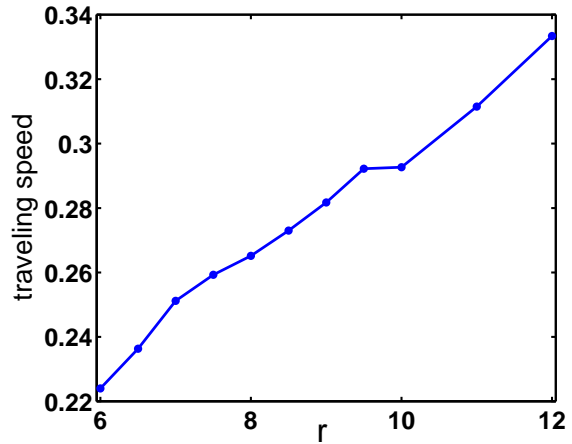}
\caption{\label{speed} (Color online) Variation of traveling speed for different values of gradient coupling strength $r$ for $\epsilon=0.6.$      }
\end{figure}

 We have discussed earlier that in the case of traveling chimera the mean phase velocity is not a suitable characteristic measure. An alternative way to distinguish the static and traveling chimera is to identify the location of the neurons on a circular ring. An effective  tool to identify the traveling chimera and its traveling speed has been proposed by \citet{scholleco}. According to their argument, in traveling chimera, each neuron stays part of the time in coherent domain and part of the time in incoherent domain without changing their spatial shape, so the maximum values of each nodes $x_i(t)$ will also change with time. For this purpose, we calculate the maximum values of $x_i(t)$ over a long time interval as $M(t)=\mbox{max}\{x_1(t), x_2(t), \cdots, x_N(t)\}$. Similarly, one can consider the minimum values of $x_i(t)$. Figure~\ref{power}(a) shows the variation of maximum values of $x_i(t)$ for a long time interval at $r=8.0$. From this figure the time series of $M(t)$ shows a periodic behavior thus indicating the episodic occurrence of maxima to a certain node. The Fourier transform of $M(t)$ determines the time period $T$, which is required for a single node to attain the maximum amplitude state. The Fourier transform of $M(t)$ is shown in Fig.~\ref{power}(b). We find that the maximum amplitude occurs at frequency $f=0.00133$ with long period $T=1/f=751.88$. With this time period $T$, the entire incoherent and coherent domains cover the whole circular ring consisting of $N$ nodes with traveling speed =$N/T=0.266$. At lower value of $r$, e.g. at $r=6.0$, the variation of $M(t)$ is shown in Fig.~\ref{power}(c) and the corresponding Fourier transform is shown in Fig.~\ref{power}(d). Here, it is found that the maximum amplitude occurs at frequency $f=0.00112$ with time period $T=892.86$. At $r=6.0$, the maximum amplitude occurs at lower frequency than that of $r=8.0$ [see inset figure of Fig.~\ref{power}(d)]. 
Figure~\ref{speed} shows the variation of traveling speed for different values of $r$. The speed of traveling chimera states increases for increasing  value of gradient coupling strength $r$.

\begin{figure}
\centering
\includegraphics[width=0.49\textwidth]{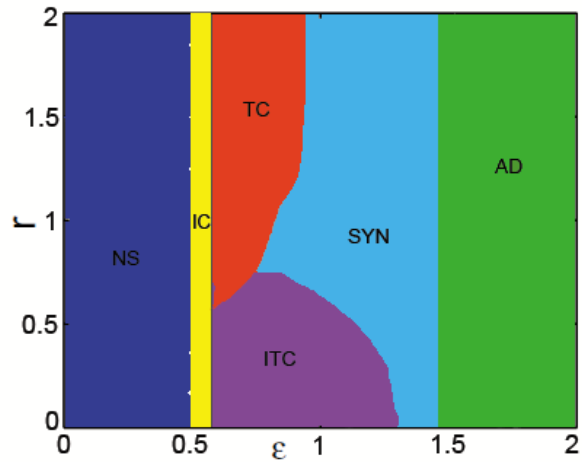}
\caption{\label{F:phase} (Color online) Phase diagram in the $\epsilon-r$ plane. NS: turbulence or unsynchronized state, IC: imperfect chimera, ITC: imperfect traveling chimera, TC: traveling chimera,  SYN: global synchronized state, AD: global amplitude death.}
\end{figure}

In order to explore the complete spatiotemporal dynamics of the system we compute the phase diagram in the $\epsilon-r$ parameter space [Fig.~\ref{F:phase}] for the range of $\epsilon \in [0, 2]$ and $r \in[0, 2]$. We change both $\epsilon$ and $r$ with step size of 0.01.  From the phase diagram it can be noticed that the choice of $\epsilon$ and $r$ organizes the phase diagram. Below a value of $\epsilon$ ($\epsilon\lesssim0.5$), the network shows turbulent behavior irrespective of the value of $r$. Beyond that we notice imperfect chimera for a very narrow region, but after this region the system shows imperfect traveling chimera for $\epsilon>r$, and traveling chimera for $\epsilon \le r$. It can be noticed from the phase diagram (Fig.~\ref{F:phase}) that for $\epsilon=r$ the system enters into the global synchrony for much lower value of $\epsilon$. For a larger value of $\epsilon$ all the neurons in the network eventually attain a stable homogeneous steady state and thus global amplitude death occurs in the network. In the parameter space, the separation lines of turbulence to imperfect chimera, imperfect chimera to traveling chimera and synchronization to amplitude death are almost vertical, which means the value of synaptic coupling strength $\epsilon$ plays an important role in determining such states, while the gradient coupling strength $r$ plays a crucial role in organizing the transition from the (imperfect) traveling chimera state to the synchronized state.
\section{Conclusion}

In this paper we have reported the occurrence of several chimera patterns in a network of Hindmarsh-Rose neuronal oscillators, which are connected by local chemical synaptic gradient coupling. Remarkably, we have found chimera state even in one-way local coupling, which reduces the essential connectivity requirement to observe chimera below nearest neighbor local coupling. Depending upon the gradient coupling parameter we observe imperfect traveling chimera, imperfect chimera and traveling chimera, of which the imperfect traveling chimera has not been observed earlier for any other coupling scheme. We have observed the imperfect traveling chimera for asymmetric excitatory coupling. Whereas for the one-way excitatory coupling and also for simultaneous excitatory-inhibitory coupling we have observed traveling chimera: although the direction of travel in those states are reversed. We also observed pronounced change in the time series of the neuronal oscillators: for predominating excitation, square-wave bursting changes to plateau bursting and then amplitude death occurs; in contrast, when both inhibitory and excitatory coupling are present, with the increase of gradient coupling strength ($r$) square-wave bursting eventually changes into the spiking behavior.

Since here we have considered a nearest neighbor and one-way local coupling thus it is intuitive to observe traveling wave solutions but it is {\it counter-intuitive} to observe chimera patterns. The present study is important in the context of neuroscience in the sense that this study gives evidence that local synaptic (gradient) coupling not only provides synchrony \cite{igor1} or amplitude death \cite{dhamala} but under certain circumstances it may drive the neuronal network to chimera state.

\begin{acknowledgments}
T.B. acknowledges financial support from SERB, Department of Science and Technology (DST), India (Grant No. SB/FTP/PS-005/2013).
\end{acknowledgments}



\providecommand{\noopsort}[1]{}\providecommand{\singleletter}[1]{#1}%
\end{document}